\documentclass{elsarticle}
\usepackage[margin=2.5cm]{geometry}
\usepackage{amsmath,amsbsy,amssymb,amsthm,mathtools,bm}
\usepackage{euscript,mathrsfs}
\usepackage{graphicx}
\usepackage{wrapfig}
\usepackage[dvipsnames]{xcolor}
\usepackage{wasysym}
\usepackage{verbatim}
\usepackage{empheq}
\usepackage{adjustbox}
\usepackage{xcolor}
\usepackage{mathrsfs} 
\usepackage{hyperref}
\definecolor{urlcolor}{HTML}{990000}
\definecolor{linkcolor}{HTML}{005F5F} 
\hypersetup{pdfstartview=FitH,  linkcolor=linkcolor,urlcolor=urlcolor, colorlinks=true,citecolor=blue}
\usepackage{tikz}
\usetikzlibrary{positioning}
\usetikzlibrary{calc}

\definecolor{mylightred}{RGB}{211,79,73}
\definecolor{mydarkred}{RGB}{199,44,38}
\definecolor{mylightgreen}{RGB}{78,153,67}
\definecolor{mydarkgreen}{RGB}{43,129,33}
\definecolor{mylightpurple}{RGB}{150,107,178}
\definecolor{mydarkpurple}{RGB}{126,78,160}
\definecolor{mylightblue}{RGB}{49,101,205}
\definecolor{mydarkblue}{RGB}{20,92,205}

\tikzset{
  juliadot/.style args={#1,#2}{shape=circle,line width=0.03ex,minimum width=0.4ex,fill=#1,draw=#2}
}

\newcommand\julialetter[1]{{\strut\fontfamily{cmss}\bfseries\selectfont{#1}}}

\DeclareRobustCommand\julia{%
\begin{tikzpicture}[baseline=0mm, every node/.style={inner sep=0mm, outer sep=0mm}]
\node[anchor=base]        (j) at (0,0) {\julialetter{\j}};
\node[anchor=base, right=0ex of j] (u) {\julialetter{u}};
\node[anchor=base, right=0ex of u] (l) {\julialetter{l}};
\node[anchor=base, right=0ex of l] (i) {\julialetter{\i}};
\node[anchor=base, right=0ex of i] (a) {\julialetter{a}};
\path let \p1 = (j) in node[juliadot={mylightblue,mydarkblue}] (bluedot) at (\x1+0.02ex,1.4ex) {};
\path let \p1 = (i) in node[juliadot={mylightred,mydarkred}] (reddot) at (\x1,1.4ex) {};
\path let \p1 = (reddot) in node[juliadot={mylightpurple,mydarkpurple}] (purpledot) at (\x1+0.5ex,\y1) {};
\path let \p1 = (reddot) in node[juliadot={mylightgreen,mydarkgreen}] (greendot) at (\x1+0.25ex,\y1+0.42ex) {};
\end{tikzpicture}%
}

\definecolor{urlcolor}{HTML}{120099}
\definecolor{linkcolor}{HTML}{005F5F}
\hypersetup{pdfstartview=FitH,  linkcolor=linkcolor,urlcolor=urlcolor, colorlinks=true,citecolor=blue}

\renewcommand{\phi}{\varphi}

\DeclareMathOperator{\re}{\mathrm{Re}\,}

\begin{document}

\title{Penrose method for Kuramoto model with inertia and noise}

\author[add1,add2]{Artem Alexandrov}
\ead{aleksandrov.aa@phystech.edu}
\author[add2,add3]{Alexander Gorsky}
\address[add1]{Moscow Institute of Physics and Technology, Dolgoprudny 141700, Russia}
\address[add2]{Institute for Information Transmission Problems, Moscow, 127994, Russia}
\address[add3]{Laboratory of Complex Networks, Center for Neurophysics and Neuromorphic Technologies, Moscow, Russia}

\begin{abstract}
    Using the Penrose method of instability analysis, we consider the synchronization transition in the Kuramoto model with inertia and noise with all-to-all couplings. Analyzing the Penrose curves, we identify the appearance of cluster and chimera states in the presence of noise. We observe that noise can destroy chimera and biclusters states. The critical coupling describing bifurcation from incoherent to coherent state is found analytically. To confirm our propositions based on the Penrose method, we perform numerical simulations.
\end{abstract}

\maketitle

\tableofcontents

\section{Introduction}

Synchronization phenomenon is widely observed in nature \cite{Pikovsky2001}, which provoking a natural desire to investigate in details the essence of it. Kuramoto model, proposed in \cite{Kuramoto1975,Kuramoto1984}, is one of the many models of synchronization and attracts a great attention due to both simplicity and existence of non-trivial physics. Despite huge progress in the investigation of Kuramoto model properties, some blank spaces still exist. These white spots have physical motivation and deserve a formulation in rigorous mathematical form. Since much of the research devoted to the Kuramoto model uses numerical simulations, we believe that the new analytic results are interesting and eligible.

In this paper, we use quite new results concerning the Kuramoto model on weighted graphs with a large number of nodes, obtained by Medvedev, Chiba \& Mizuhara in the series of papers \cite{Chiba2019P1,Chiba2019P2,Medvedev2022,Chiba2022,Chiba2023}. The cornerstone of their research is the so-called Penrose method \cite{Penrose1960}, which easily allows to capture bifurcations and transitions between different states, appeared in the model. We apply the proposed method for the Kuramoto model on a weighted graph with inertia and noise and discuss how the presence of noise affects synchronization, appearance of phase structures (patterns), and bifurcations related to transitions between different structures. Our goal is twofold. First, we want to generalize the results obtained with the help of the Penrose method. Second, we want to understand how the noise affects synchronization transition, appearance of chimeras \& clusters states, and hysteresis, caused by inertia term. The main passage of this work is that dissipative, inertial, and noise terms are naturally interconnected to each other and should be considered together. Armed with the Penrose method, we can treat all the actors (inertia, noise, dissipation) simultaneously, easily switch between different limiting cases (like zero noise limit or vanishing inertia term) and incorporate graph structure as well.

The rest of the paper is organized as follows. We start from~\autoref{sec:SetUp} with a discussion of previous research, emphasizing the development of methods, and establishing the role of our findings. In~\autoref{sec:MixingStability} we apply the Penrose method for a model with inertia and noise and then provide the explicit simple form of the expression for critical coupling constant. Developing these ideas, we analyze the impact of noise on different structures that appear in the model in~\autoref{sec:NoiseImpactPatterns}. Next, we reinforce our results by numerical simulation in~\autoref{sec:Numerics}. We finalize the narration in~\autoref{sec:DiscussConcl}. We provide the details of derivations in the Appendix.

\section{Set-up and model properties}
\label{sec:SetUp}

We consider the Kuramoto model with inertia and noise. The equations of motion are given by
\begin{equation}\label{eq:2nd-SDE-EOM}
    m\ddot{\phi}_i + \gamma \dot{\phi}_i =  \omega_i + \frac{2K}{N}\sum_{j=1}^{N}W_{ij}\sin\left(\phi_j-\phi_i\right) + \alpha_i,
\end{equation}
where $m$ is the mass (inertia) of $i$-th oscillator, $\gamma$ is dissipative factor, $\omega_i$ is the oscillator eigenfrequency, $K$ is the coupling constant, $N$ denotes the number of oscillators, $\alpha_i$ represents the white noise, $\langle \alpha_i(t)\alpha_j(t')\rangle=2D\delta_{ij}\delta(t-t')$. We start from a model defined on a weighted symmetric graph with adjacency matrix $W_{ij}$ but mainly focus on the case of complete graph. The synchronization is measured by the order parameter,
\begin{equation}
    h(t) = \frac{1}{N}\sum_{i,j=1}^{N}W_{ij}\exp\left\{i\phi_j(t)\right\}.
\end{equation}
Virtually always, the parameters of the model~\eqref{eq:2nd-SDE-EOM} are rescaled. Let us carefully consider the rescaling procedure. First, we note that $[m]=T^1$, i.e. mass term has dimensionality of time. For other parameters, we see that $[\gamma]=T^{0}$, $[\omega_i]=T^{-1}$, $[K]=T^{-1}$, $[D]=T^{-1}$. The structure of the equations tells us that parameters $m$ and $\gamma$ linked to each other and moreover if we want to rescale equations, the damping factor $\gamma$ intertwines to eigenfrequencies and coupling constant. The crucial fact is that the damping factor is dimensionless and we can set $\gamma=1$. Equations of motion~\eqref{eq:2nd-SDE-EOM} can be represented as
\begin{equation}\label{eq:1st-SDE-EOM}
\begin{aligned}
    d\phi_i &= \omega_i\,dt + v_i\,dt+\alpha_i\,dt,\\
    dv_i &= -\frac{v_i}{m}\,dt + \frac{1}{m}\frac{2K}{N}\sum_{j=1}^{N}W_{ij}\sin\left(\phi_j-\phi_i\right)\,dt
\end{aligned}
\end{equation}
Note that original set-up of Kuramoto model with noise, proposed by Sakaguchi in~\cite{Sakaguchi1988} contains (as should be) stochastic term only on phases, whereas in case with inertia the noise usually is related to velocities, i.e. $\alpha_i(t)$ appears in the second line of~\eqref{eq:1st-SDE-EOM}. The limiting cases of the proposed model have been considered in many papers. Let us briefly discuss them for the sake of completeness. The overdamped model, i.e. with vanishing inertia term, was considered rigorously by Lancelotti \cite{Lancellotti2005}, who has proven that dynamics of such model converges in $N\rightarrow\infty$ limit to Vlasov equation (in case of zero noise) or to Fokker-Planck equation (in the presence of noise). First results in stability analysis of the corresponding equations were obtained by Mirollo and Strogatz \cite{Strogatz1991}. Next, for our best knowledge, the consideration of model with inertia was started by Tanaka and coauthors in the paper \cite{Tanaka1997}. In this research, the authors have applied ideas developed by Levi with coauthors in the work \cite{Levi1978}. The crucial point of the model with inertia is the existence of hysteresis, i.e. synchronization and desynchronization occurs at different coupling constants. The underlying mechanism of such phenomenon is a coexistence of stable fixed point and running periodic solution. Discussing the model with inertia and noise, we should keep in mind the existence of hysteresis.

Moving onto the continuum limit with inertia, Acebron first discussed the limit $N\rightarrow\infty$ for the model with noise and inertia in \cite{Acebron2000}. The further development was done by Olmi et. al \cite{Olmi2014} (model with inertia) and Gupta et. al \cite{Gupta2014} (model with inertia and noise). In parallel to this research, the rigorous results were obtained by Chiba in \cite{Chiba2015} for model without inertia and by Dietert in the paper \cite{Dietert2016}, where the Penrose method for the Kuramoto model was introduced.

As was shown in paper \cite{Gupta2014}, there are several well-known limits of the proposed model. First, the overdamped limit corresponds to the Kuramoto model with noise (Kuramoto-Sakaguchi model) \cite{Sakaguchi1988}. Second, overdamped limit with identical frequencies coincides with so-called Brownian mean-field (BMF) model. Third, limit with vanishing dissipation and identical eigenfrequencies corresponds to the Hamiltonian mean field (HMF) model \cite{Dauxois2002}. So, in case of complete graph (all-to-all couplings) we have three parameters: distribution of eigenfrequencies, amplitude of noise, and ratio between particle mass and damping factor.

In this paper, we focus on the continuum limit of the model~\eqref{eq:1st-SDE-EOM}. For the conventional Kuramoto model, the consideration of continuum limit allows to easily observe low-dimensional dynamics. This dynamics is related to the invariance under M\"{o}bius group transformations and in case of finite $N$ such an invariance is the cornerstone of the Watanabe-Strogatz (W-S) ansatz \cite{Watanabe1993,Watanabe1994,Marvel2009}. In case of $N\rightarrow\infty$, the Ott-Antonsen (O-A) ansatz, proposed in \cite{Ott2008}, captures this invariance via the reduction of an infinite hierarchy of differential equations for moments of distribution function into one equation. Such a reduction corresponds to the existence of an attracting invariant manifold in the space of probability measures, the so-called O-A manifold. It is worth mentioning that the O-A manifold can be not attracting \cite{Engelbrecht2020}. The O-A ansatz fails in the presence of noise, but the expansion in terms of cumulants provides insight for systematic treatment of the case of non-zero noise \cite{Goldobin2019}. The cumulant expansion also is one of the candidates to develop the O-A ansatz in the presence of inertia term. Another way to investigate the case with inertia is the so-called adiabatic elimination, which works for small inertia. The summary of attempts to establish O-A ansatz for the model with inertia and noise is given in \cite{Permyakova2021}.

Discussed generalizations of Kuramoto model alongside approaches (W-S ansatz, O-A ansatz, cumulant expansion, etc.) are mostly focused on the case of all-to-all coupling between degrees of freedom. A general theory for more complicated architectures of couplings is quite a hot topic. Among different perspectives, the fresh view on this problem was done by Medvedev, Chiba and Mizuhara in papers \cite{Chiba2019P1,Chiba2019P2} and then developed in \cite{Medvedev2022,Chiba2022,Chiba2023}. In these papers, the authors use the concept of graphons in graph theory followed by application of the Penrose method. Based on these results, we state that the model~\eqref{eq:2nd-SDE-EOM} dynamics converges in $N\rightarrow\infty$ limit to the Fokker-Planck (F-P) equation. It means that on the phase space of model eq.~\eqref{eq:1st-SDE-EOM} in $N\rightarrow\infty$ limit there exists a density function $f=f(t,\phi,v,\omega)$ and the empirical measure obtained from eq.~\eqref{eq:1st-SDE-EOM} converges almost surely to the measure computed with function $f$. The F-P equation for $f$ is given by
\begin{equation}\label{eq:F-P-eq}
    \frac{\partial f}{\partial t}+\frac{\partial}{\partial\phi}\left\{\left(v+\omega\right)f\right\}+\frac{\partial}{\partial v}\left\{\left(-v+\mathcal{N}[f]\right)f\right\}-D\frac{\partial^2f}{\partial\phi^2}=0,
\end{equation}
where we set $m=1$ and the quantity $\mathcal{N}[f]$ is given by
\begin{equation}
    \mathcal{N}[f]=\frac{K}{i}\left\{h(t,x)e^{-i\phi}-\overline{h(t,x)}e^{+i\phi}\right\},
\end{equation}
and $h=h(t,x)$ is local continuum version of the order parameter, which is sensitive to the structure of graph via its graphon limit,
\begin{equation}
    h(t,x)=\int_{-\pi}^{+\pi}d\phi\int_{-\infty}^{+\infty}dv\int_{-\infty}^{+\infty}d\omega\int_{0}^{1}dy\,W(x,y)f(t,\phi,v,\omega,y)e^{i\phi}.
\end{equation}
Here $W=W(x,y)$ is the graphon of graph with adjacency matrix $W_{ij}$. The graphon is treated as a limit of graph with an infinite number of vertices \cite{Lovasz2012}. The function $W(x,y)$ is a symmetric function, defined on unit square $[0,1]\times[0,1]$ and this function raises the integral operator,
\begin{equation}
    \bm{W}[f](\bullet)=\int_{0}^{1}dy\,W(\bullet,y)f(y)
\end{equation}
which acts as $\bm{W}:L^2(I)\rightarrow L^2(I)$, $I\equiv[0,1]$. This is a compact self-adjoint operator with a countable sequence of eigenvalues and with a single accumulation point at zero. The equation~\eqref{eq:F-P-eq} plays the key role in our investigation. The first question that we ask is the following: how does the presence of noise affect the stability of the incoherent state?

\section{Stability of mixing state}
\label{sec:MixingStability}

\subsection{Linearized Fokker-Planck equation}

We call the function $f_0=\delta(v)g(\omega)/2\pi$ a \emph{mixing state}. It solves the Fokker-Planck equation~\eqref{eq:F-P-eq} and corresponds to incoherent state, i.e. state with order parameter equals to zero. We analyze stability of this incoherent state against of small perturbations. To do this, we perform standard derivations. First, we introduce the Fourier expansion of the distribution function $f$,
\begin{equation}\label{eq:Fourier-expansion}
    u_j(t,\zeta,\eta,x)=\int_{-\pi}^{+\pi}d\theta\,\int_{-\infty}^{+\infty}d\omega\,\int_{-\infty}^{+\infty}dv\,e^{i(j\theta+\zeta v+\eta\omega)}f(t,\theta,v,\omega,x),\quad j\in\mathbb{Z},\,\zeta\in\mathbb{R},\,\eta\in\mathbb{R}.
\end{equation}
In terms of Fourier modes, the Fokker-Planck equation~\eqref{eq:F-P-eq} becomes
\begin{equation}\label{eq:F-P-Fourier}
    \frac{\partial u_j}{\partial t}=\left(j-\zeta\right)\frac{\partial u_j}{\partial\zeta} + j\frac{\partial u_j}{\partial\eta}+K\zeta\left\{hu_{j-1}-\overline{h}u_{j+1}\right\}+Dj^2u_j,
\end{equation}
Following the paper \cite{Chiba2022}, we introduce the change of variables,
\begin{equation}
    \zeta - j =\begin{cases}-e^{-\xi_j},\quad \zeta-j<0, \\ +e^{-\xi_j},\quad \zeta-j\geq 0.\end{cases}
\end{equation}
which gives us,
\begin{equation}\label{eq:F-P-Fourier-new-vars}
    \frac{\partial u_j}{\partial t}=\frac{\partial u_j}{\partial\xi_j}+j\frac{\partial u_j}{\partial\eta}+K\left(j-e^{-\xi_j}\right)\left\{hu_{j-1}-\overline{h}u_{j+1}\right\}+Dj^2u_j,
\end{equation}
where the functions $u_j$ now depend on $\xi_j$ (we do not introduce new variable in order to be clearer). Now, we are ready to investigate the stability of the mixing state. To do it, we linearize the eq.~\eqref{eq:F-P-Fourier-new-vars} near the mixing state. The linearized version of the Fokker-Planck equation becomes (see \cite{Chiba2022} for the detailed derivation),
\begin{equation}
    \frac{\partial u_j}{\partial t}=\frac{\partial u_j}{\partial\xi_j}+j\frac{\partial u_j}{\partial\eta}+K\left(j-e^{-\xi_j}\right)\left\{hu_{j-1}-\overline{h}u_{j+1}\right\}+Dj^2u_j.
\end{equation}
Next, we represent the linearized equation as
\begin{equation}
    \frac{\partial w_1}{\partial t}=\frac{\partial w_1}{\partial\xi_1}+\frac{\partial w_1}{\partial\eta}+K\left(1-e^{-\xi_1}\right)\left\{h\hat{g}(\eta)+hw_0-\overline{h}w_2\right\}+Dw_1,
\end{equation}
where $\hat{g}(\eta)$ is the Fourier image of $g(\omega)$, $w_0 = u_0-\hat{g}(\eta)$ and $w_j=u_j$ for $j\neq 0$. In equivalent form we can write
\begin{equation}
    \frac{\partial w_1}{\partial t}=\bm{L}_1[w_1]+K\bm{B}[w_1]=\bm{S}[w_1],\quad 
    \frac{\partial w_j}{\partial t}=\bm{L}_j[w_j],\quad j\geq 0,\,j\neq 1,
\end{equation}
where we have introduced the following operators,
\begin{equation}
\begin{gathered}
    \bm{L}_j[w](\xi,\eta,x)=\left(\frac{\partial}{\partial\xi}+j\frac{\partial}{\partial\eta}+Dj^2\right)w(\xi,\eta,x),\\
    \bm{B}[w](\xi,\eta,x)=\left(1-e^{-\xi}\right)\hat{g}(\eta)\bm{W}[w(0,0,\bullet)](x).
\end{gathered}
\end{equation}
To proceed further, we need to compute the resolvent of the operator $\bm{L}_j$, which is done in the Appendix. Having found the resolvent, we can reformulate the linear stability problem as the eigenvalue problem for the linear operator.

\subsection{Eigenvalue problem}

We consider the eigenvalue problem,
\begin{equation}
    \lambda w = \bm{S}w=\left(\bm{L}_1+K\bm{B}\right)w,
\end{equation}
where the operators $\bm{L}_1$ and $\bm{B}$ are
\begin{equation}
    \bm{L}_1=\left(\partial_{\xi}+\partial_{\eta}-D\right),\quad \bm{B}[w]=\left(1-e^{-\xi}\right)\hat{g}(\eta)\bm{W}[w(0,0,\bullet)](x)
\end{equation}
and $\hat{g}(\eta)$ is the Fourier image of $g(\omega)$. The eigenvalue problem can be rewritten as
\begin{equation}
    \left(\lambda-\bm{L}_1\right)w=K\bm{B}w\rightarrow w=K\left(\lambda-\bm{L}\right)^{-1}\left(1-e^{-\xi}\right)\hat{g}(\eta)\bm{W}[w(0,0,\bullet)].
\end{equation}
Using the expression for the resolvent, we find
\begin{equation}
    \left(\lambda-\bm{L}_1\right)^{-1}\left(1-e^{-\xi}\right)\hat{g}(\eta)=\int d\omega\,g(\omega)e^{i\eta\omega}\left(\frac{1}{\lambda-i\omega+D}-\frac{e^{-\xi}}{\lambda+1-i\omega+D}\right).
\end{equation}
Next, we define the function $\mathcal{D}(\lambda,\xi,\eta)$,
\begin{equation}
    \mathcal{D}(\lambda,\xi,\eta)=\int d\omega\,g(\omega)e^{i\eta\omega}\left(\frac{1}{\lambda-i\omega+D}-\frac{e^{-\xi}}{\lambda+1-i\omega+D}\right).
\end{equation}
With this definition, the eigenvalue problem is written down in the more compact way,
\begin{equation}
    w=K\mathcal{D}(\lambda,\xi,\eta)\bm{W}[w(0,0,\bullet)].
\end{equation}
Setting $\xi=\eta=0$, we obtain the following,
\begin{equation}
    w(0,0,x)=K\mathcal{D}(\lambda,0,0)\bm{W}[w(0,0,\bullet)].
\end{equation}
Let $\mu>0$ be an eigenvalue of the operator $\bm{W}$ and $V$ be the corresponding eigenfunction. Let $G(\lambda)=\mathcal{D}(\lambda,0,0)$. Then a root of the equation $G(\lambda)=(K\mu)^{-1}$ determines the eigenvalue $\lambda=\lambda(\mu)$ as a function of $\mu$. If this holds, then $w(0,0,x)=V(x)$, which means that the set of roots of equation $G(\lambda)=(K\mu)^{-1}$ coincides with the eigenvalues of the operator $\bm{S}$. Next, the equation $G(\lambda)=(K\mu)^{-1}$ has a root if the curve $\mathcal{C}=G(it)$, $t\in\mathbb{R}$ has a positive winding number about $(K\mu)^{-1}$. The curve $\mathcal{C}$ is called critical curve or Penrose curve. The analysis of stability based on the curve $\mathcal{C}$ and its winding numbers is called Penrose method. We see that the effect of the network structure is encoded in the spectrum of the operator $\bm{W}$. Its spectrum is directly related to the spectrum of the linear operator $\bm{S}$, which is responsible for the stability of the mixing state.

\subsection{Penrose method in presence of noise}

\begin{figure}
    \centering
    \includegraphics[width=\linewidth]{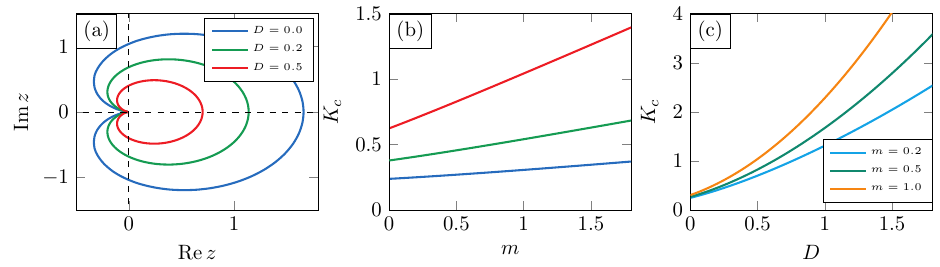}
    \caption{(a) Penrose curves for Gaussian distribution with $\sigma=0.3$ for different white noise amplitudes $D$, (b) critical coupling $K_c$ as function of $m$, (c) $K_c$ as function of $D$ for different $m$}
    \label{fig:PCurves_D}
\end{figure}

Now we present how the mentioned above Penrose method allows to determine critical coupling $K_c$ for the case of complete graph (details concerning the Penrose method are provided in Appendix). In such case, the operator $\bm{W}$ has only one eigenvalue $\mu=1$. The critical curve $\mathcal{C}=G(it)$, $\in\mathbb{R}$ can be represented in parametric form,
\begin{equation}
\begin{gathered}
    x(t)=\int_{-\infty}^{+\infty}\frac{d\omega\,g(\omega)D}{D^2+(\omega-t)^2}-m\int_{-\infty}^{+\infty}\frac{d\omega\,g(\omega)\left(1+mD\right)}{\left(1+mD\right)^2+m^2\left(\omega-t\right)^2}, \\ 
    y(t)=\int_{-\infty}^{+\infty}\frac{d\omega\,g(\omega)(\omega-t)}{D^2+(\omega-t)^2}-m^2\int_{-\infty}^{+\infty}\frac{d\omega\,g(\omega)(\omega-t)}{\left(1+mD\right)^2+m^2\left(\omega-t\right)^2}.
\end{gathered}
\end{equation}
We have plotted Penrose curves for different parameters in fig.~\ref{fig:PCurves_D}a.
Analysis of Penrose curve allows to easily find the critical value of coupling constant $K$. As we have stated above, the spectrum of operator $\bm{S}$ depends on the winding number of Penrose curve about $K^{-1}$. In case of a symmetric distribution $g(\omega)$, the curve $\mathcal{C}$ is symmetric about the real axis. It intersects the real axis at a unique point $x_0>0$. Furthermore, $G(0)=x_0$. It means that if $K<K_c$, the point $K^{-1}$ lies outside the Penrose curve. For $K>K_c$, the point lies inside the curve and the winding number is positive. Therefore, $K_c^{-1}=x_0$ and $x_0=x(0)$. Taking into account this fact, we obtain the explicit expression for the critical coupling constant,
\begin{equation}\label{eq:K-critical}
    K_c^{-1}=D\int_{-\infty}^{+\infty}\frac{d\omega\,g(\omega)}{D^2+\omega^2}-m\left(1+mD\right)\int_{-\infty}^{+\infty}\frac{d\omega\,g(\omega)}{(1+mD)^2+m^2\omega^2}.
\end{equation}
This expression is our main result. It can also be presented in the slightly different form
\begin{equation}\label{eq:K-critical-rescaled}
    K_c^{-1}=\int_{-\infty}^{+\infty}\frac{d\omega\,g(\omega D)}{1+\omega^2}-\int_{-\infty}^{+\infty}
    \frac{d\omega}{1+\omega^2}g\left(\omega D\left(1+\frac{1}{mD}\right)\right)
\end{equation}
The first term  corresponds to the limit of $m\rightarrow 0$ and it is the well-known critical coupling for the Kuramoto model with noise. The second term contains both inertia and noise contributions, but its structure is the same. In the limit $D\rightarrow 0$ the critical coupling for the model with inertia is reproduced. A slightly more general limit $D\rightarrow 0$, $m\rightarrow\infty$ with $mD=\mathrm{const}$ can be considered as well. The dependence of $K_c$ on parameter is monotonic both for $m$ and $D$, see fig.\ref{fig:PCurves_D}b and fig.\ref{fig:PCurves_D}c 

Note that this expression is dissimilar to the critical coupling obtained in \cite{Acebron2000}. The reason lies in that the authors consider a different setup with noise coupled to velocities of particles, instead of phases. In addition, they deal with another incoherent state, which can depend on velocities.

\section{Noise impact on model patterns}
\label{sec:NoiseImpactPatterns}

It is known that the Kuramoto model has a quite rich set of states, distinct from the incoherent \& coherent states. Among them, chimeras and synchronized clusters attract lot of attention. One of the possible mechanisms to realize the formation of such structures is to consider the bimodal (two-humped) frequency distribution $g(\omega)$. In this setup chimeras and clusters appear via the unfolding of mixing state bifurcation, which was discussed in details in \cite{Medvedev2022}. Noise is an additional ingredient, which can affect the mentioned patterns. In this section we use the formalism of Penrose curves in order to understand the impact of noise on different structures observed in the model with inertia.

\subsection{Bicluster states}

To investigate the bicluster state, we consider bimodal Gaussian distribution,
\begin{equation}
    g(\omega;\mu,\sigma_1,\sigma_2)=\frac{1}{2\sqrt{2\pi}}\left[\frac{e^{-(x+\mu)^2/(2\sigma_1^2)}}{\sigma_1}+\frac{e^{-(x-\mu)^2/(2\sigma_2^2)}}{\sigma_2}\right]
\end{equation}
Our reasoning is that for Gaussian distribution we can immediately compare our results with previously obtained in \cite{Chiba2023} for zero noise. Here we consider symmetric bimodal distribution, i.e. $\sigma_1=\sigma_2=\sigma$.

We fix $\sigma=0.3$ and $m=1.0$. Guided by the previous research, we know that for large enough value of $\mu$ the bicluster state arises via the Andronov-Hopf (A-H) bifurcation. In our case the appearance of such state depends on noise magnitude $D$. In terms of Penrose curve, at large enough value $\mu_c$, the curve develops a cusp, so the inverse function theorem does not work. This corresponds to the A-H bifurcation and occurs when $dy/dt|_{t=0}=0$. To find the critical value $\mu_c$, starting from that A-H bifurcation takes place, we compute the function $J(\mu)=\left.dy/dt\right|_{t=0}$. The critical value $\mu_c$ corresponds to a root of the equation $J(\mu_c)=0$.
\begin{figure}
    \centering
    \includegraphics{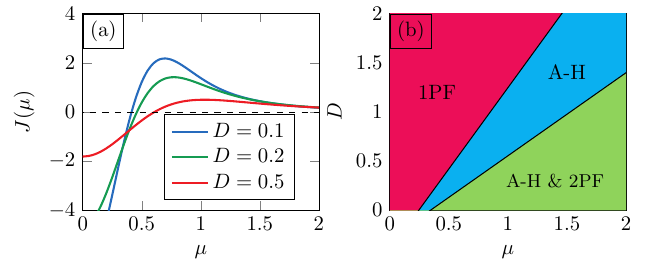}
    \caption{(a) function $J=J(\mu)$ with $m=1$ for bimodal Gaussian distribution with $\sigma=0.3$, (b) bifurcation domains in $(\mu,D)$-space}
    \label{fig:J_mu_Curves}
\end{figure}
We observe that the noise gradually increases the critical value $\mu_c$ (see fig.~\ref{fig:J_mu_Curves}a) and it means that large noise prevents the formation of bicluster structure. There is a line in the $(\mu,D)$-plane that separates the domain with the A-H bifurcation from the domain with first pitchfork bifurcation (1PF) only, which is shown at fig.~\ref{fig:J_mu_Curves}b. 

\begin{figure}
    \centering
    \includegraphics{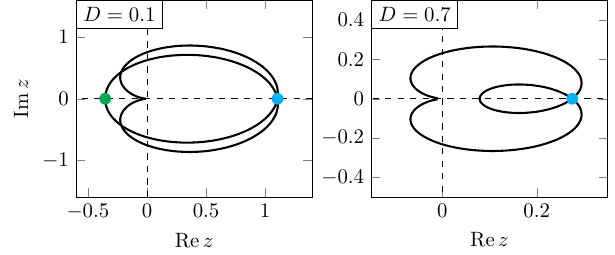}
    \caption{Disapperance of the second pitchfork bifurcation (parameters are $m=1.0$, bimodal Gaussian distribution with $\mu=1.0$, $\sigma=0.3$ is considered). Blue dot represents the point $x_0$ of the Andronov-Hopf, green dot represents the second pitchfork bifurcation}
    \label{fig:2ndPicthFork_Disappear}
\end{figure}

In addition to the A-H bifurcation, the second pitchfork bifurcation (2PF) occurs at a certain value $\mu=\mu_0$. It represents the transition from the mixing state to anti-phase coherent clusters. Noise affects such a transition in the expected way: a large enough value of the noise magnitude can shift the corresponding point $x_0$ of the Penrose curve from the negative semiaxis to the positive one. For the parameters from fig.~\ref{fig:2ndPicthFork_Disappear} this occurs at $D\approx 0.55$. At $D\approx 0.55$ the point $x_0$ crosses the line $\text{Im}\,z=0$, which results in the disappearance of the second pitchfork bifurcation. So, we conclude that in the red-colored domain on fig.~\ref{fig:J_mu_Curves} the Penrose curves are diffeomorphic to the curves shown at fig.\ref{fig:PCurves_D}. In the domain colored by blue the Penrose curves are diffeomorphic to curve shown at fig.~\ref{fig:2ndPicthFork_Disappear} for $D=0.7$, whereas in the green filled domain curves are diffeomorphic to curve with $D=0.1$ from the same figure.

\subsection{Chimeras}

\begin{figure}[h!]
    \centering
    \includegraphics[width=\linewidth]{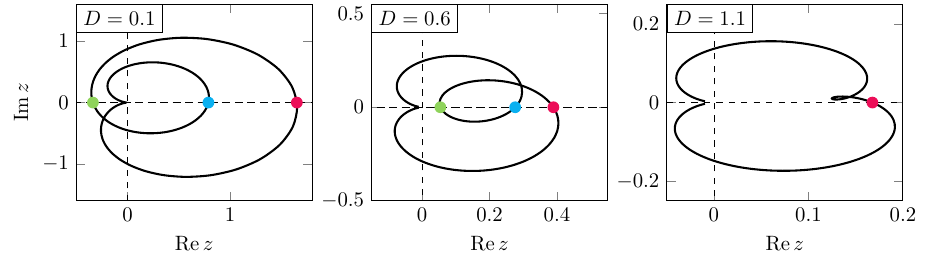}
    \caption{Destruction of chimera states by increasing of noise magnitude $D$}
    \label{fig:PCurves_Chimeras_D}
\end{figure}
Now we break the symmetry of bimodal Gaussian distribution by setting different values of $\sigma_1$ and $\sigma_2$. This causes the appearance of chimera states, i.e. the presence of synchronized and nonsynchronized oscillators in the whole population. Again, noise has a considerable influence on the chimera states. To illustrate this statement, we choose $m=1$, $\sigma_1=0.4$, $\sigma_2=0.2$, and $\mu=1$. Having fixed these parameters, we vary the noise magnitude $D$ from zero to non-zero values. The critical curves are shown at fig.~\ref{fig:PCurves_Chimeras_D}. We see that the first bifurcation (red dot) occurs at quite small value of coupling constant, and this corresponds to the partial synchronization of population, but a sizeable fraction of oscillators remains non-synchronized. The second bifurcation (blue dot) occurs at a larger value of the coupling constant, and this corresponds to the disappearance of the chimera state: now the whole population is synchronized.

\begin{figure}
    \centering
    \includegraphics[width=\linewidth]{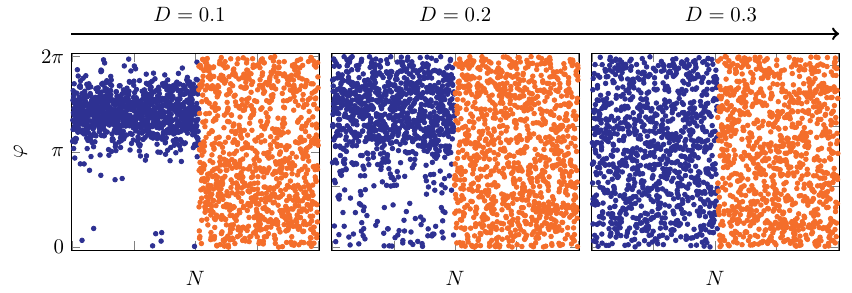}
    \caption{Destruction of chimera state by increasing noise magnitude $D$ ($N=2000$ oscillators, $K=1.0$, bimodal Gaussian distribution with parameters $\mu=1.0$, $\sigma_1=0.4$, $\sigma_2=0.2$, oscillators are grouped by intrinsic frequency sign)}
    \label{fig:Gaussian_Chimera_1}
\end{figure}

\begin{figure}
    \centering
    \includegraphics[width=\linewidth]{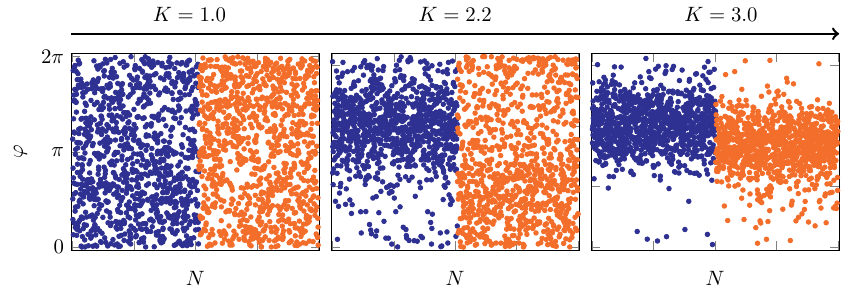}
    \caption{Bifurcations of mixed state in presence of noise for increasing coupling constant $K$ ($N=2000$ oscillators, $D=0.4$, bimodal Gaussian distribution with parameters $\mu=1.0$, $\sigma_1=0.4$, $\sigma_2=0.2$, oscillators are grouped by intrinsic frequency sign)}
    \label{fig:Gaussian_Chimera_2}
\end{figure}

\begin{figure}[h!]
    \centering
    \includegraphics[width=\linewidth]{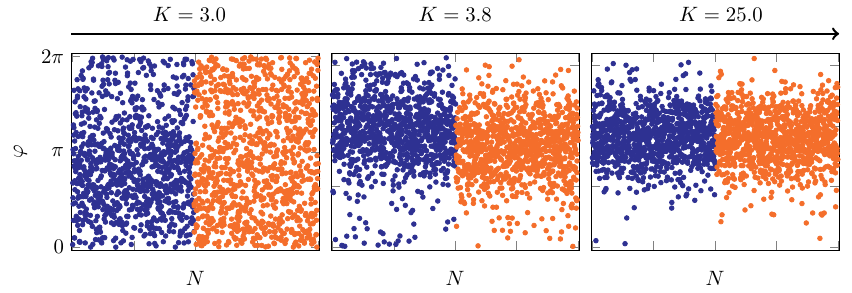}
    \caption{Bifurcations of mixed state in presence of noise for increasing coupling constant $K$ ($N=2000$ oscillators, $D=0.6$, bimodal Gaussian distribution with parameters $\mu=1.0$, $\sigma_1=0.4$, $\sigma_2=0.2$, oscillators are grouped by intrinsic frequency sign)}
    \label{fig:Gaussian_Chimera_3}
\end{figure}

Noise makes the synchronization scenario more interesting. For small values of noise, the qualitative picture does not change, whereas from a certain value of noise \emph{one more} transition occurs (middle picture on panel, see Fig.~\ref{fig:PCurves_Chimeras_D}). Further increase of noise magnitude destroys the chimera state. The intuition based on Penrose curves treatment is confirmed by direct numerical simulations, which are given 
at fig.~\ref{fig:Gaussian_Chimera_1}, fig.~\ref{fig:Gaussian_Chimera_2}, and fig~\ref{fig:Gaussian_Chimera_3}.

Figure~\ref{fig:Gaussian_Chimera_1} represents how noise destroys the chimera state. For quite small values of noise magnitude, $D=0.1$, the chimera state survives, but the slight increase ($D=0.2$) in noise causes the decay of the synchronized fraction of the population (middle picture). Further increase of noise magnitude (to $D=0.3$) completely destructs synchronized remnants. The reverse story is depicted on fig.~\ref{fig:Gaussian_Chimera_2}. Here, we fix the moderate value of the magnitude of the noise, $D=0.4$, and observe the stability of the mixing state for different values of the coupling constant $K$. From the analysis of Penrose curves, we find two critical values, $K_c^1\approx 1.60$ (bifurcation from mixing to chimera) and $K_c^2\approx 2.48$ (from chimera to bicluster). Noise suppresses synchronization, so clusters seem friable in comparison with the noiseless case.

As we have mentioned above, for the moderate value of noise, there exist three critical points, which are determined as intersections of the Penrose curve with the real axis. The corresponding transitions are shown at fig.~\ref{fig:Gaussian_Chimera_3}. The smallest critical coupling (for our choice of parameters, it is $K_c^1\approx 2.57$) corresponds to the bifurcation of mixed state to chimera. However, moderate noise significantly suppresses synchronization (the synchronized population has order parameter $\approx 0.30$, whereas nonsynchronized has $\approx 0.05$). The next critical coupling ($K_c^2\approx 3.62$) corresponds again to the bifurcation from chimera state to bicluster state. The largest critical coupling corresponds to a merging of two clusters into a fully coherent state.

\section{Numerical simulations}
\label{sec:Numerics}

We have emphasized that the model with the inertia term exhibits hysteresis. This fact should be taken into account in numerical simulations. The details of simulations can be found in the Appendix. Phase diagrams were obtained with unimodal Cauchy distribution,
\begin{equation}
    g(\omega;\mu,\sigma)=\frac{\sigma}{\pi(\sigma^2+(\omega-\mu)^2)},
\end{equation}
and other parameters are specified in figure captions. First of all, we reproduce the previous results obtained for the model with inertia only. The key point is to see hysteresis from the phase diagram. We would also like to draw attention to the convergence of finite simulations $N$ to analytical results, derived in continuum limit $N\rightarrow\infty$. Having adopted the numerical scheme, described in \cite{Vlasov2015}, we observe the same issues as pointed out in \cite{Olmi2014}: the backward critical coupling (which describes the destruction of coherent state under the decrease of coupling constant) is almost insensitive to finite size effects, whereas the forward critical coupling (which corresponds to the appearance of the synchronized state by the increase of coupling constant) is very sensitive to finite size effects. Second, we simulate the model with inertia and noise. Again, we use the scheme mentioned above. To our surprise, the finite size effects are not so significant in the presence of noise. We observe that noise strongly suppresses the hysteresis, which is similar to the effect, observed in \cite{Gupta2014}. 

\begin{figure}
    \centering
    \includegraphics{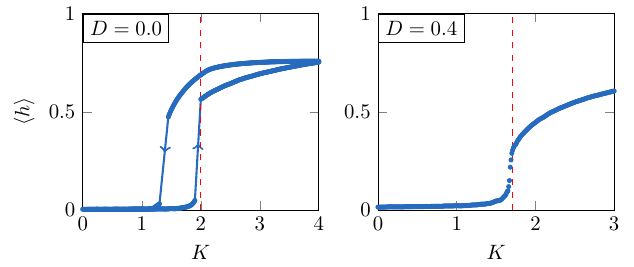}
    \caption{Left: phase diagram ($N=40000$) for unimodal Cauchy with $\sigma=1.0$, $m=1.0$ and zero noise, right: phase diagram ($N=3500$) with $\sigma=0.5$, $m=1.0$ and $D=0.4$. On both plots red line corresponds to critical coupling from~\eqref{eq:K-critical}, $\langle h\rangle$ is time-averaged order parameter.}
    \label{fig:Ph_Diag}
\end{figure}

\section{Discussion and conclusion}
\label{sec:DiscussConcl}

In this study, we consider the Kuramoto model with noise and inertia for the case of complete graph (all-to-all couplings). We analyze the stability of mixing state against of small fluctuations via linearization of the Fokker-Planck equation. The linearization procedure yields the eigenvalue problem for the particular differential operator. With help of recently obtained results, we have adopted the Penrose method to investigate the spectrum of such operator and have classified possible bifurcations of the mixing state. The formulation in terms of Penrose (critical) curves allows us to determine explicitly the critical coupling for transition to the synchronized state. The presence of noise affects on critical points corresponding to these bifurcations. It can also affect the existence of bifurcations, suppressing the transitions from mixing state to clusters (Andronov-Hopf bifurcation) and mixing state to anti-phase coherent clusters (second pitchfork bifurcation). Also, we observe that large noise destroys the chimera states.

Numerical experiments clearly show the hysteresis in the absence of noise. We have faced the known challenge of the convergence of numerics to the analytically computed critical coupling constant and observed that noise strongly affects hysteresis. We emphasize that developed in \cite{Chiba2023} method allows to capture only the so-called ``forward'' critical coupling, which corresponds to transition from non-synchronized state to the synchronized one.  Nevertheless, we assume that the Penrose method can provide a profound insight into the hysteresis phenomenon in Kuramoto model with inertia.

There are several issues for the further study. First, we have not covered the analysis of graph architecture, but it can be done in straightforward way via the graphons formalism. It should be mentioned that using of graphons is not a panacea because graphons are limits of dense graphs sequences. In more general case, so-called $P$-operators, graphop, and graphings appear (see \cite{Backhausz2022} for introduction to the topic and \cite{Gkogkas2022} for the discussion of Kuramoto model with graphop). Second, it is known that hysteresis behavior can be related to the graph architecture via frequency-degree correlated eigenfrequencies. The simplest possible example is the case of star graph, which was investigated analytically in \cite{Vlasov2015}. The analysis of synchronization on star graph with noise \cite{Alexandrov2023} shows that there is a tricritcal point, where the synchronization transition becomes continuous. For our best understanding, it is not so clear how to realize the frequency-degree correlations in terms of a graphon-based treatment. Finally, there is one more method that allows one to indicate a critical point, based on the Fisher information metric in the parameter space \cite{AlexandrovGorsky2023}. The critical points appear as singularities of this metric, defined through the stationary solutions of the Fokker-Planck equation. In the case of noise and inertia, the corresponding parameter space is three dimensional and it would be interesting to probe the hysteresis via singularities.

The usage of graphons in accompanying with the Penrose method seems to be a very powerful tool to analyze the effects of long-range interactions in the Hamiltonian mean-field (HMF) model with the additional dissipative term. The inclusion of dissipation looks natural and in the limit of vanishing dissipation all the effects known for original HMF model are reproduced. We expect that formulation in terms of graphons will allow us to tune the strength of long-range interactions, starting from the complete graph to more sparsed graphs. Introducing the dissipative term into the HMF model, we immediately relate this problem to the results for the networks of Josephson contacts and systems of granulated superconductors.

The Kuramoto model is the popular simplified model for the synchronization of neurons, see \cite{Breakspear2010} for the review. The account of noise in the neuroscience problem is standard; however, the inertia term has not been discussed in that context so far. Our study suggests that the combination of inertia and noise strongly influence at least two phenomena certainly relevant for brain synchronization: the formation and fate of chimeras \cite{Huo2023} and interplay between continuous and abrupt synchronization corresponding to the second- and first-order phase transitions. The latter is important for the transition from consciousness to unconsciousness \cite{Kim2018}.

Last point that we would like to mention is the development of neuromorphic architectures based on spintronic devices. The spin-torque oscillators (STOs) are prominent building blocks for neuromorphic computing. For instance, quite recently the set of synchronized STOs was used for vovel recognition \cite{Romera2018}. The dynamics of STO magnetization can be treated in terms of Landau–Lifshitz–Gilbert (LLG) equation, but typically this dynamics corresponds to a relatively slow magnetization change. In case of fast dynamics, memory effects should be taken into account. It can be done by introducing the inertial term into LLG \cite{Mondal2023}. The LLG equation is directly related to the dynamics of Landau-Stuart oscillators, and under some assumptions the synchronization of STOs can be treated in terms of the Kuramoto model.

\section*{Acknowledgements}

AA would like to express his huge gratitude to Georgi Medvedev for fruitful discussions and advice. Also, AA is grateful to Denis Goldobin for sharing the relevant notes and discussion of O-A ansatz for the Kuramoto model with inertia. This work is supported by the IDEAS Research Center. A.A. was supported by the Foundation for the Advancement of Theoretical Physics and Mathematics ``BASIS'' (grant №23-1-5-41-1).

\section*{Data availability}

The code snippets and the datasets generated and analyzed during the current study are available from the corresponding author on reasonable request.

\appendix

\section{Numerical simulations details}

All the numerical simulations were realized on {\julia} programming language. For a numerical solution of differential equations, we use the package \verb|DifferentialEquations|. For the simulations with nonzero noise, the Euler-Maryama was used. In case of zero noise, we have compared two implementations, based on the system of 2-nd order ODEs (via \verb|SecondOrderODEProblem|) and based on the system of 1-st order ODEs (via \verb|ODEProblem|). We observe that there are no significant differences in performance. For the computation of phase diagrams with nonzero inertia and all-to-all couplings, the following simulation scheme was used:

\begin{enumerate}
\setlength{\itemsep}{0pt}
\setlength{\parskip}{0pt}
    \item Generate initial conditions, i.e. set of $\phi_i(t=0)$. For mixing state, initial phases are drawn from uniform distribution on $(-\pi,+\pi]$
    \item Numerically solve set of $N$ differential equations for a fixed value of coupling constant $K_0$ on the interval $[0,T]$ with predefined $T$
    \item Compute time-averaged value of order parameter:
    \begin{enumerate}
    \setlength{\itemsep}{0pt}
    \setlength{\parskip}{0pt}
        \item Compute sum $h(t)=\sum_{i=j}^{N}\exp\{i\phi_j(t)\}$
        \item Compute numerically the integral $\displaystyle\langle h\rangle =\frac{1}{T}\int_{0}^{T}dt\,|h(t)|$
        \item Store $\langle h\rangle$ in memory
    \end{enumerate}
    \item Shift coupling constant by a predefined small step $\delta K$, $K_0\rightarrow K_0+\delta K$
    \item Numerically solve set of $N$ differential equations for $K_0+\delta K$ with initial conditions $\phi_i(T)$ (i.e. on this step we use the stored in memory values $\phi_i(T)$ from step 2) on the interval $[0,T]$ and refresh values $\phi_i(t=T)$ in memory
    \item Repeat steps 3, 4, 5 until the current value of coupling constant will reach predefined value $K_{\max}$ (i.e. we split interval $[0,K_{\max}]$ with predefined step $\delta K$)
\end{enumerate}

We observe quite good convergence between analytical predictions in the thermodynamic limit $N\rightarrow\infty$ and results of numerical simulations starting from $N\sim 2000$. Surprisingly, the presence of noise improves convergence. If not specified explicitly, we use $T=1000$ and time interval is discretized with step $\delta T=0.05$. The choice of $K_{\max}$ is dictated by analytical expressions for critical couplings and we set $\delta K=0.01$.

For the simulation of patterns, we find (analytically, if possible or numerically) the corresponding critical values of the coupling constant and choose the value $K^*$, which lies in an interval between the critical values (for example, $K_c^1<K^*<K_c^2$ in the case of two critical coupling constants, predicted by the Penrose curve). We generate initial conditions once and then use them. Next, we numerically solve set of $2N$ SDE and extract values of phases $\phi_i(t=T)$, $i\in\{1,N\}$. Finally, we group oscillators in accordance with the sign of their intrinsic frequencies (this is done to improve visualization, see \cite{Chiba2023}). The number of oscillators in the simulations related to patterns is $N=2000$.

\section{Operators and their spectra}

\subsection{Resolvent}

For the operator $\bm{L}_j$, we have the eigenvalue problem,
\begin{equation}
    \left(\lambda-\bm{L}_j\right)u=u.
\end{equation}
Now apply the Fourier transform $\mathcal{F}[\bullet]$ for both sides,
\begin{equation}
    \left(\lambda-i\zeta-ij\omega+Dj^2\right)\mathcal{F}[u]=\mathcal{F}[v],\,F[u](\zeta,\omega,x)=\int d\xi\,d\eta\,e^{-i\xi\zeta-i\eta\omega}u(\xi,\eta,x).
\end{equation}
It gives us $$u=\left(\lambda-\bm{L}_j\right)^{-1}u=\mathcal{F}^{-1}\mathcal{F}[u]$$. The explicit computation of the integral is
\begin{multline}
    \int d\zeta\,d\omega\,\frac{e^{+i\xi\zeta+i\omega\eta}}{\lambda-i\zeta-ij\omega+Dj^2}\int d\xi'\,d\eta'\,e^{-i\zeta\xi'-i\eta'\omega}u(\xi',\eta',x)=\\=
    \int_{0}^{\infty}dt\,e^{-\lambda t-Dj^2t}\int d\zeta\,d\omega\,e^{+i\xi\zeta+i\omega\eta}e^{+i\zeta t+ij\omega t}\int d\xi'\,d\eta'\,e^{-i\zeta\xi'-i\eta'\omega}u(\xi',\eta',x)=\\=\int_{0}^{\infty}dt\,e^{-\lambda t-Dj^2t}\int d\xi'\int d\eta'\left\{\int d\omega\,e^{i\omega(jt+\eta-\eta')}\right\}\left\{\int d\zeta\,e^{i\zeta(\xi+t-\xi')}\right\}u(\xi',\eta',x)=\\=\int_{0}^{\infty}dt\,e^{-\lambda t-Dj^2t}u(\xi+t,\eta+jt).
\end{multline}
It means that the resolvent of the operator $\bm{L}_j$ is given by
\begin{equation}
    \left(\lambda-\bm{L}_j\right)^{-1}u=\begin{cases}
    \displaystyle \int_{0}^{\infty}dt\,e^{-\lambda t-Dj^2t}u(\xi+t,\eta+jt),\quad \re\lambda>0, \\ \displaystyle -\int_{-\infty}^{0}dt\,e^{-\lambda t-Dj^2t}u(\xi+t,\eta+jt),\quad \re\lambda<0.
    \end{cases}
\end{equation}

\subsection{Eigenvalue problem}

The eigenvalue problem is given by
\begin{equation}
    \left(\lambda-\bm{L}_1\right)u=K\bm{B}u\rightarrow u=K\left(\lambda-\bm{L}\right)^{-1}\left(1-e^{-\xi}\right)\hat{g}(\eta)\bm{W}[u(0,0,\bullet)].
\end{equation}
Using the expression for resolvent, we find
\begin{multline}
    \left(\lambda-\bm{L}_1\right)^{-1}\left(1-e^{-\xi}\right)\hat{g}(\eta)=\int d\zeta\,d\omega\frac{e^{i\xi\zeta+i\eta\omega}}{\lambda-i(\zeta+\omega)+D}\mathcal{F}\left[\left(1-e^{-\xi}\right)\hat{g}(\eta)\right]=\\=\int d\zeta\,d\omega\,\frac{e^{i\zeta\xi+i\eta\omega}}{\lambda-i(\zeta+\omega)+D}\int d\xi'd\eta'e^{-i\eta'\omega-i\xi'\zeta}\left(1-e^{-\xi'}\right)\int d\omega'e^{+i\eta'\omega'}g(\omega')=\\=\int\frac{d\zeta\,d\omega\,e^{i\zeta\xi+i\eta\omega}}{\lambda-i(\zeta+\omega)+D}\int d\omega'g(\omega')\int d\xi'\left\{\int d\eta'\,e^{i\eta'(\omega'-\omega)}\right\}e^{-i\xi'\zeta}\left(1-e^{-\xi'}\right)=\\=\int d\omega\,g(\omega)e^{i\eta\omega}\left(\frac{1}{\lambda-i\omega+D}-\frac{e^{-\xi}}{\lambda+1-i\omega+D}\right).
\end{multline}

\section{Penrose method of instability analysis}

\subsection{Basic idea}

As we have stated above, the mixing state loses linear stability if the eigenvalue of operator $\bm{S}$ has a positive real part, i.e. $\re\,\lambda>0$. The eigenvalue $\lambda$ obeys the equation
\begin{equation}\label{app:disp-like-eq}
    \int_{-\infty}^{+\infty} d\omega\,g(\omega)\left(\frac{1}{\lambda-i\omega+D}-\frac{1}{\lambda+1-i\omega+D}\right) = \frac{1}{K\mu},
\end{equation}
where $\mu$ is an eigenvalue of graphon operator $\bm{W}$. For the complete graph, such eigenvalue is unique and $\mu=1$. The eq.~\eqref{app:disp-like-eq} is structurally similar to the dispersion relation, appeared in the theory of plasma and this similarity allows to apply Penrose stability criterion, originally proposed in \cite{Penrose1960}. Mapping the Penrose results onto our problem, the loss of stability corresponds to the question when the eq.~\eqref{app:disp-like-eq} has a solution with $\re\lambda>0$. To answer this question, let us consider the following function,
\begin{equation}\label{app:G-function}
    G(z)=\int_{-\infty}^{+\infty} d\omega\,g(\omega)\left(\frac{1}{z-i\omega+D}-\frac{1}{z+1-i\omega+D}\right).
\end{equation}
We represent $z=x+iy$ and take limit $x\rightarrow 0$ with help of Sokhotski-Plemelj formula, which gives us
\begin{multline}\label{app:G-x-limit}
    \lim\limits_{x\rightarrow 0}G(x+iy)=+D\int\frac{d\omega\,g(\omega)}{D^2+(\omega-y)^2}-D\int_{-\infty}^{+\infty}\frac{d\omega\,g(\omega)}{(D+1)^2+(\omega-y)^2}-\\-\int\frac{d\omega\,g(\omega)}{(D+1)^2+(\omega-y)^2}+i\int\frac{d\omega\,g(\omega)(\omega-y)}{D^2+(\omega-y)^2}-i\int\frac{d\omega\,g(\omega)(\omega-y)}{(D+1)^2+(\omega-t)^2},
\end{multline}
where all the integrals are understand in the Cauchy principle value sense. Moreover, it can be shown directly that $\lim\limits_{y\rightarrow\pm\infty}\lim\limits_{x\rightarrow 0}G(x+iy)=0$ and $G'(0)<0$ (see Lemma 3.9 in \cite{Chiba2022} for a proof). It follows that the curve $\mathcal{C}=G(it)$ is a bounded closed curve and it has an intersection with positive real axis at a unique point. Next, we apply the argument principle. It says us that the number of roots of equation $G(z)=K^{-1}$ is equal to the winding number of the curve $\mathcal{C}$ about $K^{-1}$. Therefore, we need to determine the intersection point of the Penrose curve $\mathcal{C}$ with positive real axis. Let us denote this point $x_0$. The curve $\mathcal{C}$ can be represented in parametric form and below we present corresponding derivations.

\subsection{Derivation of Penrose curve equations}

Representing $z = x+iy$, let us consider limit $x\rightarrow 0$. We denote the first integral in the definition of $G$ as $G_1$ and the second as $G_2$. So, we write $G=G_1-G_2$, where
\begin{equation}
    G_1= \int_{-\infty}^{+\infty}\frac{d\omega\,g(\omega)}{x+iy-i\omega+D},\quad G_2= \int_{-\infty}^{+\infty}\frac{d\omega\,g(\omega)}{x+iy+1-i\omega+D}.
\end{equation}
We start from $G_1$ and denote $G_1=iI_1$, where
\begin{equation}
    I_1 = \int\frac{d\omega\,g(\omega)}{ix-y+\omega+iD}=-i\int\frac{d\omega\,g(\omega)(x+D)}{(x+D)^2+(\omega-y)^2}+\int\frac{d\omega\,g(\omega)(\omega-y)}{(x+D)^2+(\omega-y)^2},
\end{equation}
and take the limit of $x\rightarrow 0$,
\begin{equation}
    \lim\limits_{x\rightarrow 0}G_1 = i\lim\limits_{x\rightarrow 0}I_1=+D\int\frac{d\omega\,g(\omega)}{D^2+(\omega-y)^2}+i\int\frac{d\omega\,g(\omega)(\omega-y)}{D^2+(\omega-y)^2}.
\end{equation}
For $G_2$, we denote $G_2=iI_2$, where
\begin{equation}
    I_2=\int\frac{d\omega\,g(\omega)}{ix-y+i+\omega+iD}=-i\int\frac{d\omega\,g(\omega)(x+D+1)}{(x+D+1)^2+(\omega-y)^2}+\int\frac{d\omega\,g(\omega)(\omega-y)}{(x+D+1)^2+(\omega-y)^2},
\end{equation}
which gives us in the limit of $x\rightarrow 0$,
\begin{equation}
    \lim\limits_{x\rightarrow 0}G_2 = i\lim\limits_{x\rightarrow 0}I_2=\int\frac{d\omega\,g(\omega)(D+1)}{(D+1)^2+(\omega-y)^2}+i\int\frac{d\omega\,g(\omega)(\omega-y)}{(D+1)^2+(\omega-y)^2}.
\end{equation}
Finally, for $G$ in the limit of $x\rightarrow 0$, we have
\begin{multline}
    G(it)=\lim\limits_{x\rightarrow 0}\left(G_1-G_2\right)=+D\int\frac{d\omega\,g(\omega)}{D^2+(\omega-t)^2}-D\int_{-\infty}^{+\infty}\frac{d\omega\,g(\omega)}{(D+1)^2+(\omega-t)^2}-\\-\int\frac{d\omega\,g(\omega)}{(D+1)^2+(\omega-t)^2}+i\int\frac{d\omega\,g(\omega)(\omega-t)}{D^2+(\omega-t)^2}-i\int\frac{d\omega\,g(\omega)(\omega-t)}{(D+1)^2+(\omega-t)^2}.
\end{multline}
It defines the curve in complex plane $(x,y)$, where $t$ is the parameter.

\bibliography{references.bib}
\end{document}